# Evidence for Incipient Ferroelectricity in YCrO$_3$

Ashish Kumar Mall[1+], Barnita Paul[1+], Ashish Garg[2], Rajeev Gupta[1, 3*]

[1]*Materials Science Programme,* [2]*Department of Materials Science and Engineering*

[3]*Department of Physics, Indian Institute of Technology Kanpur, Kanpur- 208016, India*

Cubic structure is one of the most commonly found structures for oxides at high temperature. As the temperature is lowered only a handful of these oxides exhibit ferroelectricity which is rather surprising. In this paper, we use the example of YCrO$_3$ (YCO), an incipient ferroelectric material to show that in most oxides there are two competing phenomenon – onset of ferroelectricity due to rotation of CrO$_6$ octahedra and displacement of Y atom leading to suppression of ferroelectricity. This competition reveals that while the octahedral rotations favor a lower symmetry state, the Y atom displacement opposes it leaving YCO to exhibit only an incipient ferroelectric state. These results while being in agreement with the earlier theoretical predictions can also help suggest a pathway to a more stable ferroelectric state in these oxides by using a larger cationic substitution at the Y site.

---

+ Both authors contributed equally to the work. * Corresponding author email address: guptaraj@iitk.ac.in

Ferroelectric materials have been extensively used for novel device functionalities in the field of low energy nonvolatile memory based devices [1-3]. These ferroelectric random access memory (RAM) devices are superior compared to earlier magnetic memory based devices due to its low power consumption, light weight, retention of memory even when power is interrupted and robustness to radiation [1]. In general, technologically important ferroelectrics such as $Pb(Zr_{1-x}Ti_x)O_3$ are $ABO_3$ type perovskites [4]. In such oxides the spontaneous polarization usually originates from the polar displacements of Jahn-Teller active B-site cations (commonly $d^0$ transition metals) as well as by the lone-pair active cations on the A-site [5,6]. However Van aken *et al.* have proposed another mechanism to explain the origin of spontaneous polarization known as geometric ferroelectricity in $ABO_3$ perovskites. In such systems, the ferroelectricity is induced by a structural distortion due to the cationic size effects, rather than the typical changes in chemical bonding [7]. Apart from the chemical bonding or ionic size effects, orbital ordering also plays an important role to induce ferroelectricity in transition metal oxide perovskites [8]. On the contrary, for a non-transition metal $ABO_3$ perovskites, spontaneous polarization results from a structural array of the constituent ions mediated by the cooperative displacement of B-site atoms filling the octahedral site. [4]. The $BO_6$ octahedral rotation in $ABO_3$ perovskites about one or more than one crystal axes leads to novel physical phenomena in these systems. Although oxide perovskites comprise a broad class of materials, only a very few lead free perovskite oxide ferroelectrics exist in nature [9]. To solve this puzzle, we need to understand the origin of ferroelectricity in $ABO_3$ perovskites.

An ideal $ABO_3$ perovskites possess cubic crystal structure $(Pm\bar{3}m)$. For an ideal cubic perovskite the ions are in contact with each other, that makes $(r_A + r_O) = \sqrt{2}(r_B + r_O)$, where $r_O$, $r_A$ and $r_B$ are the ionic radii of oxygen, A, and B ions. The degree of distortion in $ABO_3$ perovskite due to mismatch in cationic radii is calculated from Goldschmidt tolerance factor defined by, $t = (r_A + r_O)/\sqrt{2}(r_B + r_O)$ [10]. Perovskites with $t > 1$ include the classical ferroelectric systems, like BaTiO$_3$. Ferroelectricity in this case is believed to be B-site driven, due to the smaller B-site cationic radii (i.e., Ti). The smaller cationic radius creates displacive instability and ferroelectricity appears in the system. On the contrary, perovskites having $t < 1$ do not show ferroelectricity at all. Cations adjust the bond lengths by tilts and rotations of the BO$_6$ octahedra tending to structures like characteristic GdFeO$_3$ (*Pnma*) type structure [9]. Hence in these cases, distortions in the BO$_6$ octahedra suppress ferroelectricity in these systems. The most commonly observed unpolar space group among perovskites is *Pnma*. The tolerance factor for these perovskites lies in the range $0.8 < t < 0.9$ [11]. The orthorhombic *Pnma* structure can be derived by three different rotational distortions in ideal cubic $(Pm\bar{3}m)-$ i) an octahedral rotation about [001], ii) rotation about [110] crystallographic axis along with A-site displacement and finally iii) an anti-polar A-site displacement resulting in a small displacement of the equatorial oxygen atoms [9]. The octahedral rotation maximizes covalent bonding interactions and minimizes repulsion between cationic A-site and oxygen atom [9]. It has been shown by Benedek *et al.* [9] that A-site contribution to the ferroelectric eigenvector is comparatively larger than the contribution of B-site. Hence, coordination environment at the A-site is more unfavorable for smaller tolerance factor thus leading towards a ferroelectric distortion. Hence, it is expected that the octahedral rotation optimizes the A-site coordination and reduces the ferroelectric distortion. However, with a decrease in the tolerance factor, although tendency towards ferroelectric instability grows significantly, the A-site displacements complements octahedral rotations in the *Pnma* structure and suppresses ferroelectricity. However, all octahedral rotations do not favors the shift of A-site cation from its ideal $(Pm\bar{3}m)$ position. Woodward suggested that [12] A-site displacements provide the extra degree of freedom to optimize its coordination environment that results more desirable distribution of A−O bond lengths for the *Pnma* structure. Hence, for ABO$_3$ perovskites the A-site coordination environment is optimized by octahedral rotations as well as A site displacements in *Pnma* crystal structure.

In this context, rare-earth orthochromites (RCrO$_3$) attracts the special attention with a reasonably high ferroelectric transition temperature. However, small ferroelectricity reported for RCrO$_3$ ~ 2 μC/cm$^2$ for YCO [13] and 0.35 μC/cm$^2$ for LuCrO$_3$ [14] at room temperature despite the large A-cation off-centering distortions remains puzzling. Orthorhombic YCO possess centrosymmetric *Pnma* crystal structure. Both Y and Cr has relatively smaller ionic radii [15] yielding a tolerance factor *t* = 0.85 which is well below the critical regime where A-cation shift hardly plays a prominent role in suppressing the ferroelectricity. A recent powder neutron diffraction study by Ramesha *et al.* suggested that YCO exhibits local non centrosymmetricity near the dielectric transition ~ 440 K due to off-centering displacement of Cr$^{3+}$ ions keeping the average crystal structure centrosymmetric only. According to them, the order of displacement is ~ 0.01 Å along the *z* direction which gives rise to weak ferroelectricity ~ 2 μC/cm$^2$ in YCO [13]. Although A-site displacement could play a key role in suppressing the ferroelectricity, here the off centering of Cr$^{3+}$ cation in the B-site is the origin of local non-centrosymmetricity in the structure. Hence, we choose YCO to study the competitive role of octahedral rotation and A-cationic displacement to find the origin of suppression of ferroelectricity in this system. More recently Mahana *et al.* have shown similar polar distortion of Gd$^{3+}$ off-centering driven by local symmetry breaking in orthorhombic *Pna*2$_1$ structure of GdCrO$_3$ analogous to YCO [16]. However, both these studies have concluded the local non-centrosymmetricity based on bulk measurement techniques such as powder neutron or X-ray diffraction only. Raman spectroscopy being a local probe can provide a better insight to enquire the local distortion compared to neutron or X-ray diffraction. While most of the above experiments suggest that YCO is a room temperature ferroelectric, there are no reports of a well saturated P-E loop for this system. Therefore, it is indeed an intriguing question that even when the electrical leakage is low, why YCO is a poor ferroelectric at room temperature. An earlier report by Taniguchi *et al.* [17] have paved the way to modify macroscopic ferroelectricity in oxide perovskites by cationic substitution.

In this paper, we report the local distortion in the $CrO_6$ octahedra of centrosymmetric YCO across the paraelectric to the ferroelectric transition temperature. While XRD does not find any prominent distortion in the crystal structure with temperature, Raman measurements can probe the local non-centrosymmetricity in the structure from the anomalous phonon deviation. We believe that this study will help to understand why ferroelectric perovskites are so few in nature and will help to optimize A-site displacement with few external parameters such as cationic substitution, external pressure etc to achieve the ferroelectricity in these structures.

**Results**

1. **Absence of structural phase transition across $T_C$ from temperature dependent XRD measurements:**

We have carried out temperature dependent XRD measurements on YCO over the temperature range of 300-900 K. The XRD pattern of YCO at 300 K were indexed with the ICDD database with card no. 34-0365. Absence of any extra peak suggests a single-phase formation of YCO without any impurities present in the sample. The high grazing angle ($\omega = 4.03°$) in GIXRD was adopted to avoid the 2θ region between $30.2°–32.5°$ where a major peak of the high temperature dome material (PEEK polymer) exists. All temperature dependent XRD patterns were fitted with Rietveld refinement to find any subtle change in the crystal structure with temperature. The refined pattern shown by the black solid line matches quite well with the experimentally observed pattern shown by the red symbols. The reduced $\chi^2$ value depicts goodness of fitting for patterns at 300 K as well as at 900 K. All patterns were fitted with orthorhombic YCO phase having *Pnma* symmetry (space group 62), containing eight formula units of YCO. A schematic of *Pnma* YCO's structure is shown in the inset of Fig. 1. The corresponding atomic coordinates are Y: 4c ($x$, 0.25, $z$); Cr: 4b (0, 0, 0.5); O1: 4c ($x$, 0.25, $z$) and O2: 8d ($x$, $y$, $z$). We do not find any drastic change in the XRD pattern with increase in temperature suggesting absence of any significant structural distortion in YCO crystal structure. The orthorhombic structure of YCO corresponds to the corner sharing $CrO_6$ octahedra with Y atoms sitting at the center of the cage formed by the $CrO_6$ octahedra where the Cr ion is coordinated with apical O1 and planar O2 oxygen ions [18]. Here, Cr–O octahedron shares its corners with surrounding Y cation polyhedra, which share their edges among neighboring Cr-O octahedra. The linear change in the lattice parameters (*a, b & c*) and unit cell volume with temperature, are shown in Fig. 2 which depict a small anomaly in the linear increase of lattice parameters across para to ferroelectric transition temperature ($T_C$) ~ 460 K. The thermal expansion coefficients along three principal axes (*a, b,* and *c*) obtained from the slope of the linear fitting, as shown in Fig. 2, are $3.06\times10^{-6}$ K$^{-1}$, $7.25\times10^{-6}$ K$^{-1}$ and $7.92\times10^{-6}$ K$^{-1}$, respectively. Here, a negligible difference among these values suggests the thermal isotropic property of YCO. The thermal coefficient of volume expansion ($1.83\times10^{-5}$ K$^{-1}$) obtained from the slope of cell volume vs. temperature plot also follows the relation $\gamma = 3\alpha$, valid for isotropic materials. Further, we have estimated the bond lengths, bond angle, polyhedral volume and orthorhombic distortion parameters from the refined patterns using a formulation proposed by Momma and Izumi [19], to understand the effect of temperature on the crystal structure. We have estimated the polyhedral distortion index ($\Delta$) using an approach proposed by W. H. Baur [20]; $\Delta = \dfrac{1}{n}\sum_{j=1}^{n}\dfrac{(l_j - l_{avg})}{l_{avg}}$ where $l_j=$ distance between central atom and $j^{th}$ coordinating atom, $n$ = bond number and $l_{avg.}$ = average bond length. It is observed that, at room temperature, Cr-O octahedron exhibits least distortion ($\Delta$ = 0.0026) in comparison to Y-O polyhedra with $\Delta$ = 0.0434 and retains its original shape. However, the distortion indices of the respective polyhedra are found to increase significantly from 0.004 to 0.045 at 900 K. Orthorhombic distortion has been calculated using $\Delta D = \dfrac{a+c-b/\sqrt{2}}{a+c+b/\sqrt{2}}$. The decrement in $\Delta D$ on increasing the temperature is shown in Fig. 3. Additionally, $CrO_6$ octahedral tilt angles ($\theta$ & $\phi$), are evaluated using the refined atomic coordinates following the formalism proposed by Y. Zhao *et al.* [21]. We find a monotonic decrease in the tilt angles ($\theta$, $\varphi$) with temperature as shown in Fig. 3 with a clear anomaly across $T_C \sim 460$ K. Unlike other ferroelectrics, the observed changes in lattice parameters and its associated atomic displacements are very tiny. The refined lattice parameters a, b, and c contracted by 0.001Å, 0.002Å, and 0.001Å along the respective directions across $T_C$. This tiny change in parameters most likely represents the limit of measurements precision and confirms that the change in lattice parameters during transition is really very small. Hence, the results infer that YCO undergoes an iso-structural (orthorhombic to orthorhombic) transition with *Pnma* space group, across $T_C$.

The thermal variation of $\Delta D$, defined as the overall distortion of Y, Cr, O1 and O2 for the structural transformation, shows no signature of phase transformation except a small anomaly appear around $T_C$ is depicted in Fig. 3(b). However, it is the $CrO_6$ octahedral tilt angle ($\theta$ & $\phi$) that shows significant change, around $T_C$. This tilting of $CrO_6$ octahedra is sufficient to induce a distortion in $YO_{12}$ polyhedra that leads to *A*-cation displacement in *Pnma* structure.

## 2. Anomalous phonon deviation across $T_C$ from temperature dependent Raman measurements:

Since temperature dependent XRD measurements could not reveal any signature of a significant lattice distortion across the paraelectric to ferroelectric transition, we intend to probe the local ordering by Raman measurements. Temperature dependent Raman measurements of YCO was carried out to study the behaviour of phonons across the phase transitions. All spectra were baseline corrected with linear background and shown by + symbols in Fig. 4. Group theoretical studies yield an irreducible representation at the Brillouin zone center: $\Gamma_{Raman} = 7A_{1g} + 8A_{1u} + 5B_{1g} + 8B_{1u} + 7B_{2g} + 10B_{2u} + 5B_{3g} + 10B_{3u}$. Out of these 60 modes, 24 modes ($7A_{1g} + 5B_{1g} + 7B_{2g} + 5B_{3g}$) are Raman-active [22]. In the present study, we have observed 17 Raman active modes among 24 allowed one in the spectral range of 130–780 cm$^{-1}$ at room temperature, which matches fairly well with the previous studies on YCO [22–24]. The higher Raman lines are assigned to the vibrations of oxygen atoms (O1 and O2) only whereas the lower Raman lines are attributed to the vibrations of both Y and O atoms [22]. The most intense and sharp Raman peak observed at 219 cm$^{-1}$ is related to the Y-cationic motion in *x*-direction, whereas the peaks at 267 cm$^{-1}$, 281 cm$^{-1}$, 409 cm$^{-1}$ and 426 cm$^{-1}$ correspond to the CrO$_6$ octahedral rotation. Two peaks appearing at 486 and 568 cm$^{-1}$ are relatively weak in intensity, and corresponds to the Cr–O stretching modes [22].

Two selected spectral ranges fitted with the sum of Lorentzian line shapes represent the temperature dependent Raman spectra between 300 K to 600 K. Red solid lines show the fitted spectra. The spectral features change due to expected thermal broadening with increase in temperature. We do not find appearance or disappearance of any new mode, as shown in Fig. 4, revealing absence of any structural change across the anomaly regime in the vicinity of $T_C \sim 460$ K. This also signifies that the crystal structure remains unchanged over the para to ferroelectric phase transition. The evolution of peak positions and line widths for the selected modes of YCO spectra, are plotted in Fig. 5 as a function of temperature. The peak positions and the line width show red shift and blue shift upon increasing the temperature, respectively. Such behavior is expected because of the lattice expansion and an increase in the phonon population with increasing temperature. In addition, we have observed an anomalous change in the line shape parameters with temperature in the observed twelve intense Raman modes near the $T_C$ ~ 460 K as depicted in Fig. 5. It is expected that Raman modes would shift to lower wave-numbers with increase in the temperature due to softening of the bond length. However in our case, Raman peak positions vs temperature, as shown in Fig. 5, show that the modes from 300 K to 465 K follow expected anharmonicity. However, there is a clear anomaly over 465 K – 490 K following expected anharmonic behavior beyond 490 K up to 600 K. It is to be recalled that refined lattice parameters (a, b, c) also show a minute deviation across $T_C$ from the otherwise increasing trend, as shown in Fig. 2. Hence the anomalous hardening of modes can be explained due to change in the lattice parameters. Interestingly, earlier reports on the temperature dependent permittivity data [25] for YCO have also reported such anomaly across $T_C$. These anomalies are much more pronounced in line widths vs temperature plot, as depicted in Fig. 5. Although the line-width increases as expected due to thermal broadening, there is a sudden change in the line widths across $T_C$. Since Raman line-width is inversely proportional to phonon life-time, we believe that the local distortion in $CrO_6$ octahedra around $T_C$ decreases the phonon lifetime and increases the line width by a magnitude much larger than what is expected due to thermal broadening only.

### 3.    Softening of phonon modes around T$_C$ :

From the above discussion, it appears that some phonon modes show strong anharmonicity across T$_C$. Since lattice vibrations are collective movements of ions, soft mode frequency estimated from the freezing of a strongly anharmonic lattice vibration, can conventionally be used as an order parameter of ferroelectricity. Hence, in this section we look carefully into the softening behavior of some specific phonon modes to explain the incipient ferroelectricity in YCO. Unlike many other conventional ferroelectric materials (e.g. BaTiO$_3$, SrTiO$_3$ and LaAlO$_3$) [26], unusually large number of modes in YCO show an anomalous change across T$_C$. In YCO, no mode softens to zero frequency across the paraelectric to ferroelectric phase transition. The number of modes seen across the transition remain the same suggesting that the system undergoes an isostructural phase transition and a permanent dipole moment is induced due to the Cr-cation displacement in the octahedral cage. In such systems, the soft mode frequency acts as an order parameter and diminishes with lowering of temperature. For these systems, the mode frequency decreases anomalously as the transition temperature is reached and is best described by the relation: $\omega(T) = \omega_0 \times (T_C - T)^\beta$ where $\beta$ is the critical exponent and line-width fitted with, $\Gamma(T) = \Gamma_0 \times (1 - T/T_C)^\gamma$ where $\gamma$ is another critical exponent [26]. It should be mentioned that the exponents $\beta$ and $\gamma$ take the value of 0.5 for a conventional phonon soft mode. In our Raman spectra, the peak frequency vs temperature of the Raman modes did not show any conventional soft mode like behaviour as seen in SrTiO$_3$ [26]. Although, almost all phonon modes show anomalous behavior in peak position and the line-width across T$_C$, we focus on the specific modes near 219, 342, 492, and 565 cm$^{-1}$ that show maximum deviation from the expected anharmonicity. It is worthwhile to mention that these modes also correspond to the atomic motion involving A-cationic displacement and rotation on CrO$_6$ octahedra. Hence, the temperature dependent behavior of these modes will provide a better understanding in the role of octahedral rotation and A-cation displacement in the origin of improper ferroelectricity in the system. The schematic vibrational patterns of these modes are shown in Fig. 6 (a)-(d). Since none of these modes behave like conventional soft mode, we consider that $(\omega - \omega_c)^n \approx (T - T_C)$. To find the scaling law of these modes with temperature around $T_C$, we have plotted $\ln|\omega - \omega_C|$ and $\ln|\Gamma - \Gamma_C|$ vs T in Fig. 6 (e)-(f). It should be noted that unlike conventional soft modes, here $\omega^2$ does not remain zero below $T_C$ for any of the above mentioned modes suggesting local distortion is present only for temperatures close to $T_C$. Hence, we have subtracted the critical value of peak position and line-width at $T_C$ i.e. $\omega_c$ and $\Gamma_c$ respectively. The solid line shows the linear scaling of peak-position and line-width for all the above-mentioned modes. The slope of the linear fit for above modes vary from 0.44±0.03–0.54±0.03. From the inverse of the slope of the linear fit, we estimate $n$ to be close to 2, the value one would expect for a conventional soft mode from mean-field approximation. On the other hand, the slope for line-width appears to be 0.26±0.02 − 0.34±0.03. The close value of slope for all above-mentioned modes prove that phonons modes involving octahedral rotation and A-cation shift behave in a similar way and that these motions cooperate each other. In addition, we have found that this holds true for all other modes also (not shown here). It is interesting to see that, contrary to the conventional soft phonon modes, the modes shown in our case only soften to zero as they approach $T_C$. They recover again as soon as the temperature deviates from the $T_C$ *i.e.* 465 K.

**Discussion:**

From the above results it is observed that YCO behaves like an incipient ferroelectric material. The suppression of ferroelectricity in this compound can be understood from the following discussions. Below $T_C$, competitive interplay between the ferroelectric distortion and octahedral rotations in ABO$_3$ materials to stabilize the crystal structure can be written by expanding the free energy ($F$) about the cubic $Pm\overline{3}m$ phase up to fourth order which can be expressed as [9]:

$$F = F_P + F_R + F_{PR}.......(1)$$

where, $F_P$, $F_R$, $F_{PR}$ represent the free energies corresponding to ferroelectric distortion, octahedral rotation separately and the coupling between them. Expanding the ferroelectric and rotational distortion about cubic $Pm\overline{3}m$ phase up to fourth order free energy term can be expressed as $F_P \sim \alpha_P Q_P^2 + \beta_P Q_P^4$, $F_R \sim \alpha_R Q_R^2 + \beta_R Q_R^4$ and the coupled term $F_{PR} \sim \gamma_{PR} Q_P^2 Q_R^2$, where, $Q$ represents the amplitudes of modes and $\alpha$, $\beta$, $\gamma$ are the corresponding coefficients. The suffixes represent the distortion corresponding to ferroelectric distortion, rotation and coupled terms respectively. Hence the frequency of the ferroelectric mode ($\omega_P$) can be written as $\omega_P^2 \sim \alpha_P + \gamma_{PR} Q_R^2$. For coupling coefficient $\gamma_{PR} < 0$, $\omega_P^2$ becomes more negative enhancing the ferroelectricity aided by octahedral rotations. On the contrary, for $\gamma_{PR} > 0$, $\omega_P^2$ increases suggesting suppression of ferroelectricity due to octahedral rotation. Although rotations can completely suppress the ferroelectricity in compounds having larger tolerance factor e.g. GdScO$_3$, it never happens for lower tolerance factor compounds. Rather, it is the A-site displacements that play a subtle role to completely suppress ferroelectricity in many *Pnma* materials with smaller tolerance factors [9].

To find the correlation between the A-site cationic displacement and the octahedral rotation leading to suppression of ferroelectricity, we look carefully into the temperature evolution of phonon modes involving specifically corresponding atomic vibration. From Fig. 6 we have seen that the phonon modes involving CrO$_6$ octahedral rotation and A-cationic displacements, the phonon modes do not soften to zero below T$_C$. We believe that if the system involves CrO$_6$ octahedral rotation only without any A-cationic displacements, $\omega^2$ would remain zero. However, it is the A-cationic displacements that play a crucial role to renormalize the phonon frequency and prohibit the softening of modes to be zero like conventional soft modes. From the scaling of phonon modes shown in Fig. 6(e)-(f), it can be assumed that the $CrO_6$ octahedral rotation behaves like a cooperative phenomenon and it affects the adjacent octahedral cage or unit cell in a similar way. Since Raman scattering probes only the zone center phonons, these modes would not then scale in the same manner with almost equal slope as shown in Fig. 6(e-f). The fact that $Cr^{3+}$ is a non Jahn-Teller ion also supports the cooperative rotation of $CrO_6$ octahedra.

It is to be mentioned that the small size of Y makes it a good candidate to study the cooperative phenomenon of A-site displacement and $CrO_6$ octahedral rotation in suppressing the ferroelectricity. It will be interesting to see if the substitution of the Y ion with a larger atom would prevent the displacement of the A-cation, thereby causing negligible suppression of ferroelectricity. This hints towards the possibility of inducing macroscopic ferroelectricity by cation substitutions to optimize the rotations in octahedral chains in perovskite-type ferroelectrics. We believe that our study helps to find why ferroelectricity is rarely observed in the family of perovskite materials despite that octahedral rotation induces ferroelectricity in the system. To the best of our knowledge this is first experimental verification of the theoretical model provided by Benedek *et al.* [9] that A –site cationic shift indeed suppresses ferroelectricity in this *Pnma* crystal structure and hence there are so few perovskite ferroelectrics in this class of materials.

Finally in Fig. 7 we have shown a schematic phase diagram of YCO depicting phase evolution with temperature.

**Conclusions:**

In summary, we have shown the suppression of ferroelectricity in YCO associated with the A-site cationic displacements from Raman scattering. The evolution of phonon modes across $T_C$ support the role of phonons in the origin of ferroelectricity associated with the local distortion in the crystal structure. We believe that this work addresses the long-standing question as to why despite the high temperature perovskite cubic structure being so common, very few oxides transform into R3c structures and exhibit ferroelectricity. We suggest that doping of a "large" cation at A site would perhaps lead to the stabilization of the ferroelectric state in many such oxides.

**Methods**

**Chemical synthesis.** Polycrystalline sample of YCO was prepared by the conventional solid state reaction method. High purity $Y_2O_3$ (99.99%) and $Cr_2O_3$ ($\geq 98\%$) powders were mixed in a stoichiometric ratio and calcined in air atmosphere at 1200 ºC for 16 h in high density $Al_2O_3$ crucibles. Further, subsequent calcinations were done at 1300 ºC for 16 h with intermediate grindings of 1hr to attain the single phase. The final powder was compacted into pellets through hydraulic pressure and sintered at 1550 °C for 20 h in air atmosphere. The powder thus prepared was green in color.

**X-ray Diffraction.** Grazing incidence X-ray diffraction (GIXRD) was carried-out at 45 kV/40 mA power setting and a scan speed of 0.48°/min with Omega = 4.03°, to characterized the phase purity and crystal structure. Temperature dependent powder X-ray diffraction (XRD) of YCO sample was carried out using a high resolution Philips X'Pert PRO MRD diffractometer (CuK$_\alpha$ radiation; $\lambda = 1.5406$ Å) with an angular resolution of 0.0001° and step size 0.02° in the range of $20° \leq 2\theta \leq$ 75°. Domed Hot Stage (DHS 900, Anton Paar) was used as a pioneering heating attachment for high-temperature XRD study from 300K to 900K.

**Raman spectroscopy.** Micro-Raman study was carried out in back scattering geometry using an optical fiber microscope coupled to a monochromator (iHR550, Horiba, France) with Peltier cooled charge-coupled detector (Synapse, Horiba JY, France). An $Ar^+$ 514.5 nm laser line was used as an excitation source. We have used a 50 L× microscope objective lens for focusing the light on the sample. The sample temperature was varied between 300–600 K using a Linkam THMS600 stage equipped with a TMS94 temperature controller along with a liquid nitrogen pump.

**References:**

1. Scott, J. F., & Paz de Araujo, C. A. Ferroelectric memories. *Science* **246**, 1400–1405 (1989).

2. Takasu, H. The ferroelectric memory and its applications. *J. Electroceramics* **4**, 327–338, (2000).

3. Wu, S. Y. A new ferroelectric memory device, Metal-ferroelectric-semiconductor Transistor *IEEE T. Electron Dev.* **21**, 499–504 (1974).

4. Cohen, R. E., Origin of ferroelectricity in perovskite oxides. *Nature* **358**, 136–138 (1992).

5. Rondinelli, J. M., Eidelson, A. S., & Spaldin, N. A. Non-d$^0$ Mn-driven ferroelectricity in antiferromagnetic BaMnO$_3$. *Phys. Rev. B* **79**, 205119(1–6) (2009).

6. Burdett, J. K. Use of the Jahn-Teller theorem in inorganic chemistry. *Inorg. Chem* **20**, 1959 – 1962 (1981).

7. Van Aken, B. B., Palstra, T. T. M., Filippetti, A., & Spaldin, N. A. The origin of ferroelectricity in magnetoelectric YMnO$_3$. *Nat. Mater.* **3**, 164–170 (2004).

8. Keimer, B. Ferroelectricity driven by orbital order *Nat. Mater.* **5**, 933–934 (2006).

9. Benedek, N. A. & Fennie, C. J. *J. Phys. Chem. C* **117**, 13339–13349 (2013).

10. Goldschmidt, V. M. Die Gesetze der Krystallochemie. *Naturwissenschaften* **14**, 477–485 (1926).

11. Giaquinta, D. M., & H-C. zur Loye. Structural predictions in the ABO$_3$ phase diagram. *Chem. Mater.* **6**, 365–372 (1994).

12. Woodward, P. M., Octahedral tilting in perovskites. II. Structure stabilizing forces. *Acta Crystallogr. B* **53**, 44–66 (1997).

13. Ramesha, K., Llobet, A., Proffen, Th., Serrao, C. R., & Rao, C. N. R. Observation of local non-centrosymmetry in weakly biferroic YCrO₃. *J. Phys.: Condens. Matter* **19,** 102202(1–8) (2007).

14. Sahu, J. R., Serrao, C. R., Ray, N., Waghmare, U. V., & Rao, C. N. R. Rare earth chromites: a new family of multiferroics. *J. Mater. Chem.*, **17,** 42–44 (2007).

15. Shannon, R. D. Revised effective ionic radii and systematic studies of interatomic distances in halides and chaleogenides. *Acta Cryst..A* **32,** 751–767 (1976).

16. Mahana, S., Rakshit, B., Basu, R., Dhara, S., Joseph, B., Manju, U., Mahanti, S. D., & Topwal, D. Local inversion symmetry breaking and spin-phonon coupling in the perovskite GdCrO₃. *Phys. Rev. B* **96,** 104106(1–9) (2017).

17. Taniguchi, H., Soon, H. P., Shimizu, T., Moriwake, H., Shan, Y. J., & Itoh, M. Mechanism for suppression of ferroelectricity in Cd₁₋ₓCaₓTiO₃. *Phys. Rev. B* **84,** 174106(1–5) (2011).

18. Geller, S., & Wood, E. A. Crystallographic studies of perovskite-like compounds. I. rare earth orthoferrites and YFeO₃, YCrO₃, YA1O₃. *Acta Cryst.* **9,** 563–568 (1956).

19. Momma, K., & Izumi, F., VESTA: a three-dimensional visualization system for electronic and structural analysis. VESTA: a three-dimensional visualization system for electronic and structural analysis. *J. Appl. Cryst.* **41,** 653–658 (2008).

20. Baur, W., The geometry of polyhedral distortions. Predictive relationships for the phosphate group. *Acta Crystallographica B*. **30,** 1195-1215 (1974).

21. Zhao, Y., Weidner, D. J., Parise, J. B., Cox, D. E. Critical phenomena and phase transition of perovskite—data for NaMgF₃ perovskite. Part II. *Phys. Earth Planet. Inter* **76,** 17-34 (1993).

22. Iliev. M. N., Abrashev, M. V., Lee, H. –G., Popov, V. N., Sun, Y. Y., Thomsen, C., Meng, R. L., & Chu, C. W. Raman spectroscopy of orthorhombic perovskitelike $YMnO_3$ and $LaMnO_3$. *Phys. Rev. B* **57**, 2872–2877 (1998).

23. Weber, M. C., Kreisel, J., Thomas, P. A., Newton, M., Sardar, K., & Walton, R. I. Phonon Raman scattering of $RCrO_3$ perovskites (R = Y, La, Pr, Sm,Gd, Dy,Ho, Yb, Lu). *Phys. Rev. B* **85**, 054303(1–9) (2012).

24. Todorov, N. D., Abrashev, M. V., Ivanov, V. G., Tsutsumanova, G. G., Marinova, V., Wang, Y.-Q., & Iliev, M. N. Comparative Raman study of isostructural $YCrO_3$ and $YMnO_3$: Effects of structural distortions and twinning. *Phys. Rev. B* **83**, 224303(1–9) (2011).

25. Serrao, C. R., Kundu, A. K., Krupanidhi, S. B., Waghmare, U. V. &. Rao, C. N. R., Biferroic $YCrO_3$. *Phys Rev. B* **72**, 220101(R) (1–4) (2005).

26. Scott, J. F., Soft-mode spectroscopy. Experimental studies of structural phase transitions. *Rev. Mod. Phys*. **46**, 83–128 (1974).

**Figure captions:**

**Fig. 1** Rietveld refinement of powder XRD pattern of YCO sample at (a) 300 K and (b) 900 K. The solid black line represents the calculated pattern with orthorhombic *Pnma* symmetry. Green bars show the positions of Bragg reflection peaks and the blue line is the difference between the experimental and calculated patterns in each panel. Inset figure shows the schematic crystal structure of *Pnma* YCO. The color code is as follows: blue (Y), green (Cr) and red (O).

**Fig. 2** (a) Variation of lattice parameters obtained from Reitveld refined XRD patterns with temperature. The solid line represents the linear fitting. The shaded region represents the region around TC~ 470 K (b) Zoom in view of the shaded region across $T_C$.

**Fig. 3** Variation of average $CrO_6$ octahedral tilt angle (square and circle) & orthorhombic distortion (triangle) obtained from the refined XRD patterns as a function of temperature.

**Fig. 4** Temperature dependent Raman spectra from 300–600 K (all spectra are not shown) by two selected spectral ranges shown by + symbols. Red solid lines represent net fitted spectra deconvoluted with Lorentzian components.

**Fig. 5** Temperature dependent line-shape parameters (peak positions and line-width) of selected Raman modes.

**Fig. 6** Schematic vibrational pattern for (a) 219 $cm^{-1}$ (b) 342 $cm^{-1}$ (c) 491 $cm^{-1}$ and (d) 565 $cm^{-1}$. Scaling law for phonon (e) peak positions and (f) line-width of given modes. Solid lines represent the linear fitting.

**Fig. 7:** Schematic phase diagram of YCO with temperature.

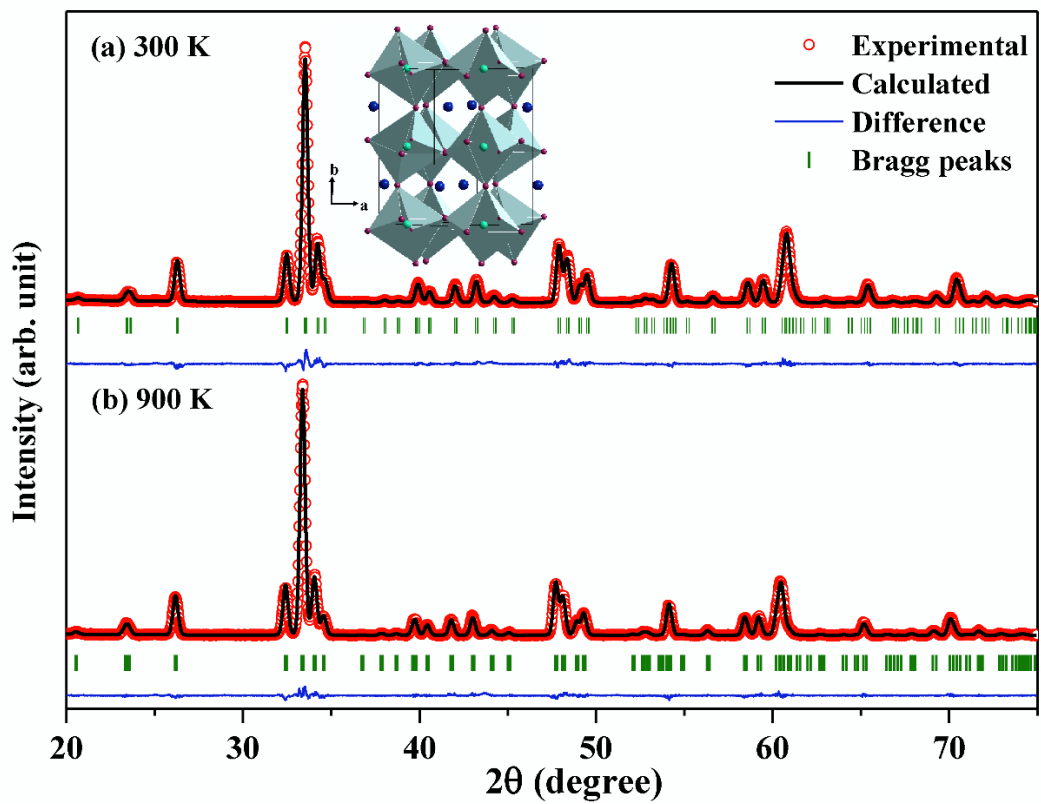

**Fig. 1**

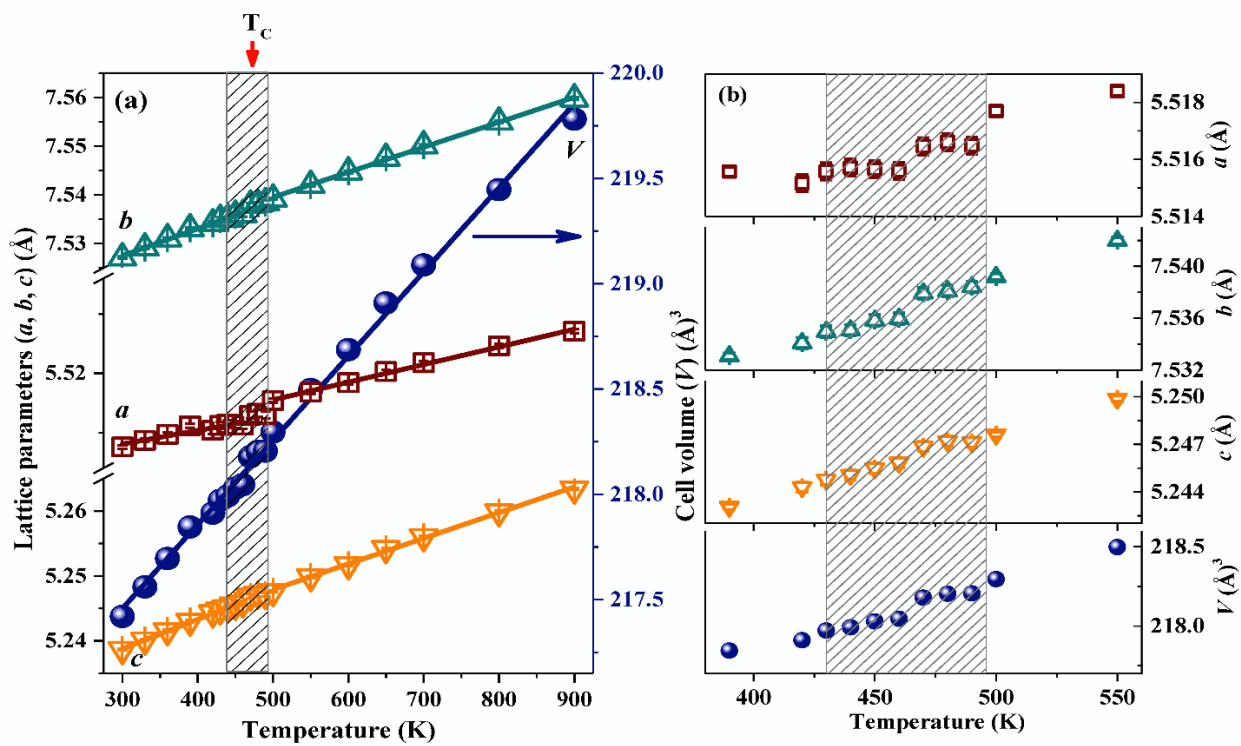

Fig. 2

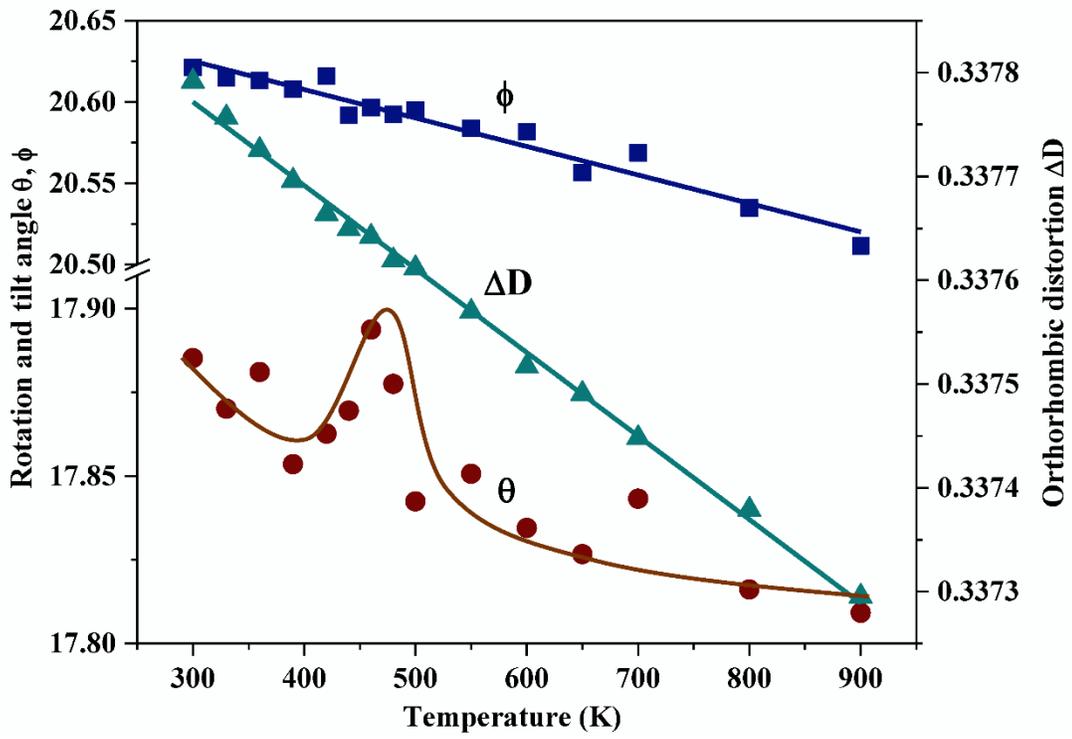

**Fig. 3**

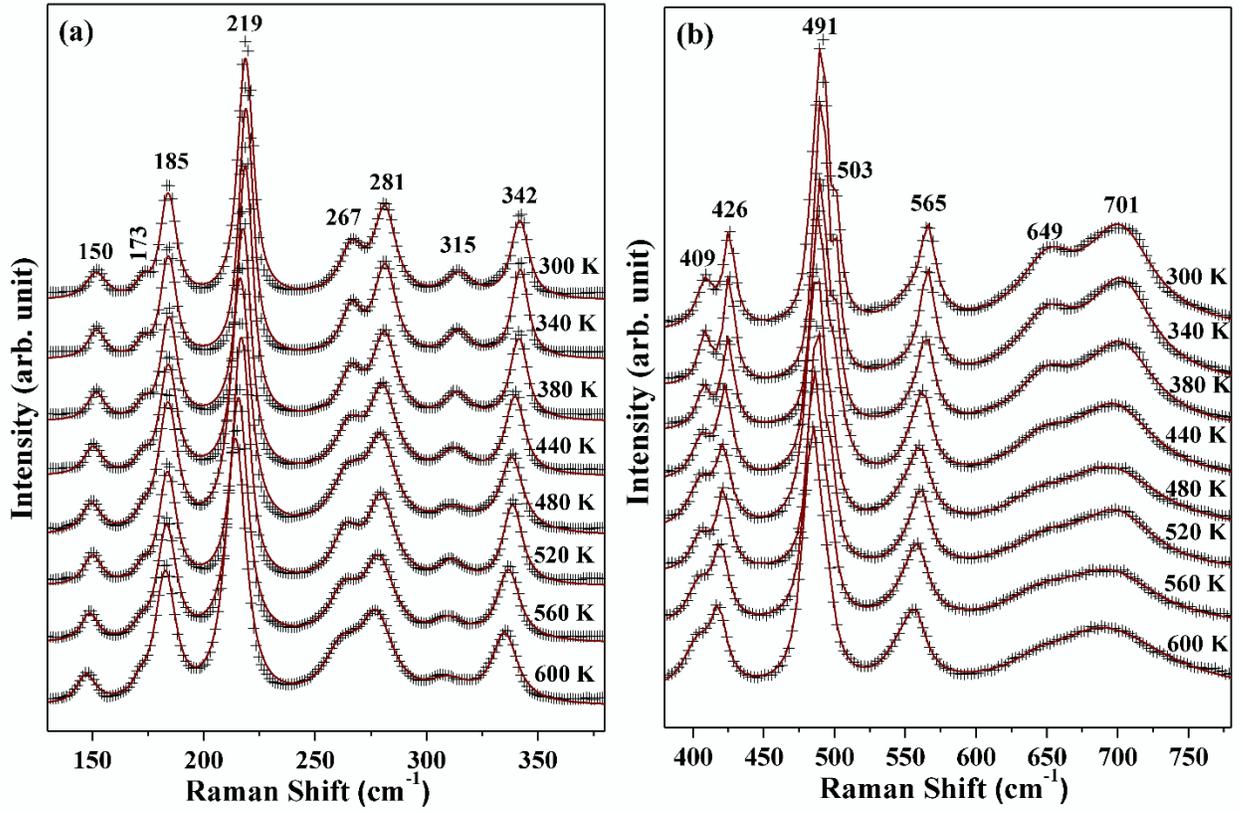

**Fig. 4**

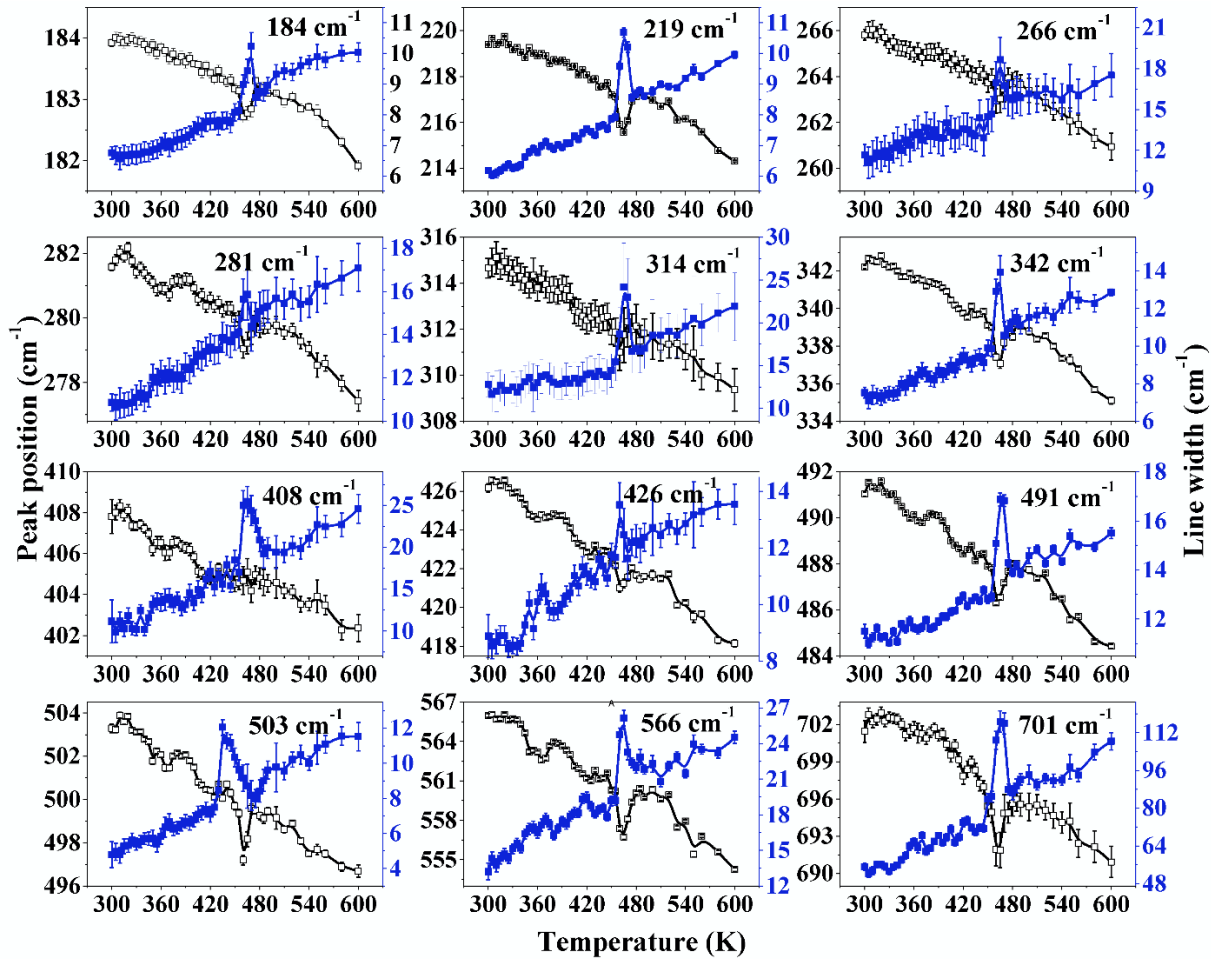

**Fig. 5**

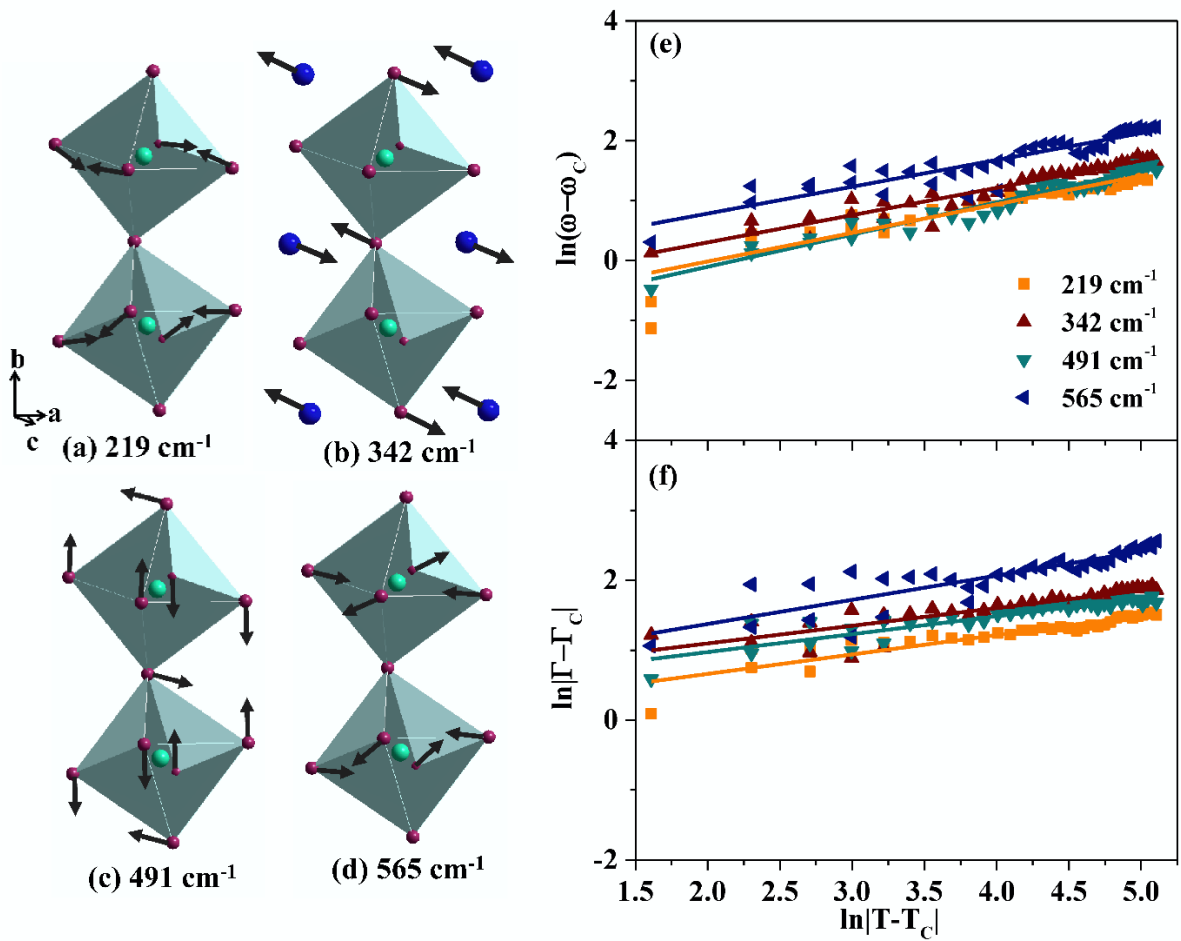

**(a)** 219 cm$^{-1}$  **(b)** 342 cm$^{-1}$

**(c)** 491 cm$^{-1}$  **(d)** 565 cm$^{-1}$

**(e)**

$\ln(\omega - \omega_C)$

- 219 cm$^{-1}$
- 342 cm$^{-1}$
- 491 cm$^{-1}$
- 565 cm$^{-1}$

**(f)**

$\ln|\Gamma - \Gamma_C|$

$\ln|T - T_C|$

**Fig. 6**

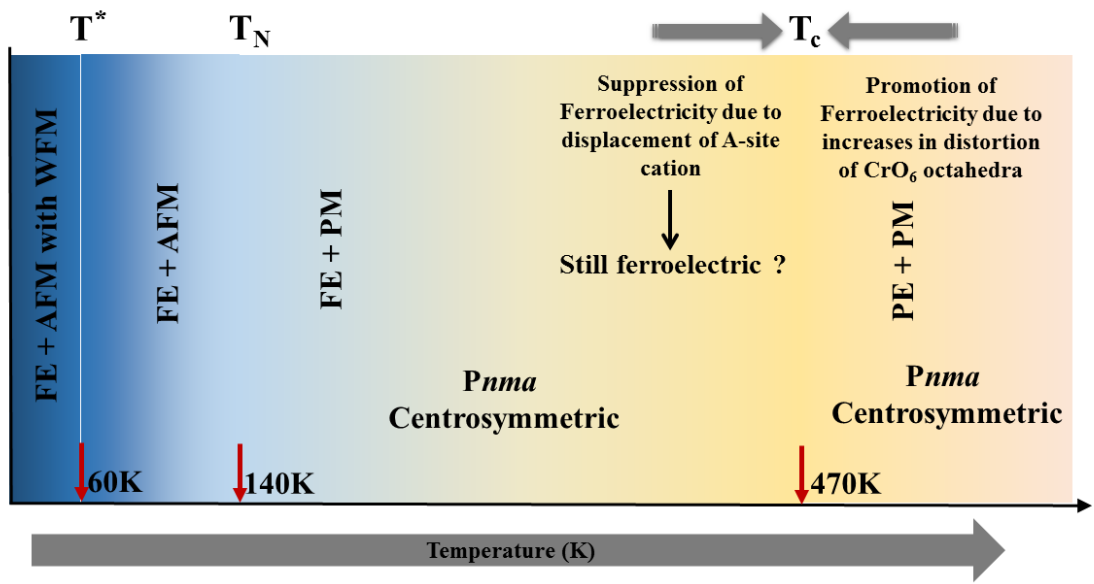

**Fig. 7**